\documentclass[12pt]{article}
\usepackage{amsmath}
\usepackage{amssymb}
\usepackage{amscd}
\usepackage[dvips]{graphicx}
 \linethickness{0.4pt}
\begin{document}

\markboth{Daszkiewicz et al.} {High spin particles with spin-mass
coupling}

\begin{center}
\section*
{High spin particles with spin-mass coupling}
\end{center}
\begin{center}
{M. DASZKIEWICZ}
\end{center}
\begin{center}
{Institute of Theoretical Physics, Wroc{\l}aw University, pl. Maxa
Borna 9\\
50-206 Wroc{\l}aw, Poland \\
marcin@ift.uni.wroc.pl}
\end{center}
\begin{center}
{Z. HASIEWICZ, C. J. WALCZYK}
\end{center}
\begin{center}
{Institute of Theoretical Physics, University of Bia{\l}ystok, ul.
Lipowa 41\\
15-424 Bia{\l}ystok, Poland \\
zhas@uwb.edu.pl, c.walczyk@alpha.uwb.edu.pl}
\end{center}

\begin{abstract}
\noindent
The classical and quantum model of high spin particles is
 analyzed  within the manifestly covariant framework. The model is obtained by supplementing the standard
Lagrange function for relativistic point particle by additional terms governing the dynamics of internal degrees of freedom.
They are described by
${\mathcal C} (3,1)$
Clifford algebra (Majorana) spinors.
     The covariant quantization
    leads to the spectrum of the particles with the masses
    depending on their spins. The particles (and anti-particles)
    appear to be orphaned as their potential anti-particle partners are of
    different mass.
\end{abstract}


\section{Introduction}
The classical and quantum model of the particle with spin dependent
mass spectrum was introduced in the article \cite{HDS}. It is
defined by the standard action for relativistic massive particle
supplemented by kinetic term for (commuting) ${\mathcal C} (3,1)$
Clifford algebra - Majorana -  spinor and covariant coupling of the
velocity with spinorial vector current. It should be stressed that
the properties of ${\mathcal C} (3,1)$ Clifford algebra are crucial
for the construction of the model. First of all, due to the fact
that  $\mathbf{SO}(3,1)$ - invariant bilinear forms on ${\mathcal C}
(3,1)$ representation space are antisymmetric,
 it is possible to construct first order kinetic terms for spinor variables - they are not total
derivatives. Secondly, the intrinsic property of bilinear spinor
"currents" being light-like is responsible for absence of fourth
order spinor terms in hamiltonian constraint. This in turn makes it
possible to define the canonical complex polarization of constraint
algebra and to parametrize the system in terms of minimal building
blocks of all $\mathbf{SL}(2,\mathbb{C}) \simeq \mathbf{Spin} (3,1)$
representations, i.e. in terms of $\mathcal{C}(3,0) \simeq {\mathcal
C} (3,1)_0$ (even subalgebra) Clifford algebra spinors.
  \\
\\
On the canonical level, the model is given as constrained system
with constraints of mixed type. By solving the second class
constraints and quantizing the resulting first class system one
obtains the description of it's spectrum in terms of Wigner basis
\cite{HDS}.

 There are
 good reasons that the presented
model is interesting from theoretical point of view. In contrast to
many other high spin particle models with additional spinorial
coordinate (see e.g. \cite{GT}-\cite{andrzej}),
    it gives quantum-mechanical, relativistic description
    of the particles with arbitrarily high spins and
    with
    spin dependent
    masses, put on Regge-like trajectory (with large masses asymptotically linear in spin).
The similar property\footnote{The similarity is not exact: the
masses of the string states
    depend directly on the string excitation levels.
    These levels are highly degenerate with respect to spin. The detailed analysis of $4$-d string content
    was presented in \cite{daszhasjas1},\cite{daszhasjas2}. }
    is shared by the string (superstring)
\cite{Pol} and field-theoretical \cite{field} models only.
    The alternative point particle model
    providing spin-dependent mass \cite{plu}, without any spinorial
    coordinates and with higher time derivatives, provides
    non-Regge behaviour of
    spin spectrum (for large spins masses go to zero). \\
    \\
     From the canonical analysis of the classical system,
     it evidently follows, that
     the crucial point of the construction presented
     here relies on the minimal coupling rule, i.e.
    on the substitution: $p^\mu \rightarrow p^\mu + j^\mu$, where
    $j^\mu$ denotes the spinorial current. This rule  seems to be consistent with the geometrical nature of the
    objects under consideration and is additionally supported by
    commonly accepted principle of minimal coupling,
   which appears to be the spin-mass coupling this time.\\
The model obtained in this way describes, at the level of the
classical equations of motion, the well known phenomenon of
"zitterbewegung" \cite{schroedinger}-\cite{rivas}. For this reason,
it can be also a starting point for the general analysis
quantum-mechanical notion of position of relativistic particles
\cite{positionoperators1}-\cite{positionoperators3} with higher
spins. These interesting questions, tightly related with fundamental
principles of quantum mechanics, are intrinsically built into Dirac
theory of spin $\frac{1}{2}$ particles with electron and
positron included as prominent examples \cite{dirac}.\\
\\
The questions raised above will not be pursued in this paper. The
main task here is to present the covariant formulation of the model
which was originally proposed in the article \cite{HDS}.\\
In contrast to earlier approach the current analysis does not rely
on solving the second class constraints but on their (complex)
polarization. The complex polarization of the classical constraints
corresponds on the quantum level to the well known
Gupta-Bleuler procedure \cite{GB}. \\
The polarized constraints give  the generalization of Dirac-type
equations (spin irreducibility) \cite{Lop} for the particles with
arbitrary spins and masses located at the  particular type of Regge
trajectory. \\
\\
The covariant formulation seems to be important from at least few
points of view. First of all, it may serve as the starting point for
the formulation of second quantized theory \cite{Jost} of high spin
particles -
with intrinsically built in spin-mass dependence.\\
    It can also be
interesting from the point of view of BRST approach \cite{brst}. In
particular, due to the presence of the second class constraints, it
enables one to investigate of  so called anomalous BRST complexes.
The objects of this kind were introduced and partially investigated
in the context of the massive string theory \cite{DHW}. It seems
that the simplicity of the particle model (finite dimensional
algebra of constraints) in comparison with the string formalism may
enable one to understand better the BRST approach to anomalous
systems.
\\
The covariant formulation can also be the starting point to the
analysis of the multiparticle system in the presence of external
fields e.g. Yang-Mills field or gravity. It is particularly
interesting to analyze the second case in the context of "graviting"
spin, as one encounters direct spin-mass dependence here. It can
appear helpful to understand the collapse and evaporation of heavy
astrophysical objects \cite{Eva}. \\
Since the masses of the particles
are fixed by (geometry dependent) kinematical constraint it would
be, by all means, interesting to investigate the model in the
presence of basic black-hole metrics e.g.
Schwarzschild, Kerr and Reissner-Nordstr\"{o}m backgrounds \cite{Schwarzschild}. \\
\\
It should be stressed that the particle spectrum obtained here is
not CPT invariant. It seems that in order to restore this symmetry
it is necessary and enough to add another spinorial degree of
freedom in an appropriate way. The full analysis of the model
modified in such a
way looks more complicated and is postponed to the future publication. \\
\\
The paper is organized as follows.\\
    In the first section the classical model is briefly recalled.
    The mixed type Poisson algebra of constraints is polarized and
    the complex Weyl coordinates are introduced.\\
The second section is devoted to the analysis of the model on the
first quantized level. The space of physical states is found by
solving the Dirac-type equations for spin irreducibility and
imposing the
kinematical constraint. \\
Finally, the results are summarized and some open questions and
problems are raised.


\section{The classical model}
The classical model considered in this paper is defined by the
following Lagrange function \cite{HDS}:
\begin{equation}
\label{action}
    {\cal L}=\frac{1}{2}e^{-1}\dot{x}^2-\frac{1}{2}e
    m_{0}^{2} +\bar{\eta}{\dot \eta} -\frac{h}{2} \dot{x}\cdot j\;.
\end{equation}
The first two terms constitute the standard action of the scalar
relativistic particle of mass $m_{0}$. It is supplemented by the
kinetic term for Majorana spinor $\eta$ and the term which couples
the particle trajectory with spinor current:
\begin{equation}
\label{current}
    j^{\mu} =  \bar{\eta}\gamma^{\mu}{\eta}\;.
\end{equation}
For the real Majorana spinors to exist it is assumed that the metric
in Minkowski space is given by: $g^{00} = -1,\,g^{ij} =
\delta^{ij}\,;\;i,j = 1,2,3$. Hence the Clifford algebra generators present in (\ref{current}) do satisfy the following
relations:
\begin{equation}
\label{clifford}
    \gamma^{\mu} \gamma^\nu + \gamma^\nu\gamma^\mu=  2 g^{\mu\nu}\;,\;\; \mu,\nu = 0,\cdots ,3\;.
\end{equation}
The Lorentz - in fact $\mathbf{Spin} (3,1)$ - invariant scalar product
of spinors  is ineviteably antisymmetric and can be explicitly
realized as
\begin{equation}
\label{convention}
    \bar{\eta}{\eta}' = {\eta}^{{\rm
    T}}\gamma^{0}{\eta}' = {\eta}^{\alpha} C_{\alpha \beta}{\eta
    '}^{\beta}\;.
\end{equation}
It should be however stressed that this explicite form will never be used in the sequel. For all the considerations in this paper it is enough
to know that the invariant form above is anti-symmetric and generates the $\beta_-$ anti-automorphism of
${\mathcal C} (3,1)$ i.e. that uniquely defined by: $\beta_-(\gamma^\mu) = - \gamma^\mu $ .
From the above one can immediately deduce
the general property of the vector current built out of
single Majorana spinor, namely that  the spinor current (\ref{current}) present in
(\ref{action}) is  light-like:
\begin{equation}
\label{null}
    j^2 = j_{\mu}j^{\mu} = 0\;.
\end{equation}
\\
The Lagrange function (\ref{action}) defines the constrained
hamiltonian system. After elimination of the canonical variables
corresponding to world-line 1-bein (e.g. by putting $e \equiv 1$),
one is left with the phase space parametrized by the particle
position and momentum $(x^{\mu},p_{\mu})$, and the canonical pairs
corresponding to the real Majorana spinor variables and their
spinorial momenta
    $(\eta^{\alpha},\pi_{\alpha} )$. Their Poisson Brackets are of standard form:
\begin{equation}
\label{poissoncan}
    \left \{p_{\mu},x^{\nu}\right \} =
\delta_{\mu}^{\nu}\;,\;\;\; \{\pi_{\alpha},\eta^{\beta}\} =
\delta_{\alpha}^{\beta}\;.
\end{equation}
The system is obviously constrained. Due to the fact that the
Lagrange function (\ref{action}) is linear in time derivative of
spinor  there are second class constraints:
\begin{equation}
\label{secondclasscon}
    G^{\alpha} = \pi^{\alpha} +
    \eta^{\alpha}\;,\;\;
     \{\,G^{\alpha},G^{\beta} \} = 2C^{\alpha
    \beta}\;.
\end{equation}
where $\pi^{\alpha} = C^{\alpha\beta}\pi_{\beta}$ and
$(C^{\alpha\beta})$ is the inverse of the matrix defined in
(\ref{convention}). \\
The above constraints are supplemented by the first class kinematic
condition which is related with the reparametrization invariance of
the action corresponding to (\ref{action}):
\begin{equation}
\label{diracham}
    H_{\rm D} = \frac{1}{2}(p^2 + m_0^2) +
    \frac{h}{2}\pi_{\alpha}p^{\alpha}_{\;\beta}\eta^{\beta}\;,\;\;{\rm
    where}\;\;p^{\alpha}_{\;\beta} = (p_{\mu}\gamma^{\mu})^{\alpha}_{\;\beta}\;.
\end{equation}
This constraint coincides with the canonical Dirac hamiltonian. One
should notice that due to (\ref{null}) the hamiltonian does
not contain the quartic terms in spinor variables.\\
The algebra of constraints is closed as in addition to
(\ref{secondclasscon}) one has:
\begin{equation}
\label{poissonmixed}
    \{H_{\rm D},G^{\alpha} \} = \frac{h}{2}p^{\alpha}_{\;\beta}G^{\beta}\;.
\end{equation}
From (\ref{secondclasscon}) and the formulae above it follows that
the constraints form the system of mixed type.\\
\\
 There are two ways of treating of the systems of this kind. One
may solve the second class constraints to obtain the first class
system on the reduced phase space. This way of proceeding was
already applied in the paper \cite{HDS}. After quantization it gave
the description of arbitrarily high spin particles in Wigner
basis. Their masses appeared correlated with spins. \\
    The other method will be applied in this paper. It gives much more
    tractable and transparent, manifestly covariant description of
the spectrum. \\
    The approach adopted below has its sources in the ideas of the
    fundamental papers of Gupta and Bleuler \cite{GB}.\\
    Instead of solving the second class constraints one may
polarize the Poisson algebra (\ref{secondclasscon}),
(\ref{poissonmixed})  to obtain an equivalent system of first class.
Due to the structure of the Poisson brackets of $H_D$ with
$G^{\alpha}$ in (\ref{poissonmixed}) the way of polarization depends
essentially on the value of $p^2$. It should be stressed that the
algebra of constraints admits the real polarization for tachionic
momenta $p^2>0$ only. The analysis of this situation is physically
less interesting and much more difficult formally. For these reasons
it will not be pursued here.
\\
 In the most interesting case $p^2<0$, which corresponds to the (real)
massive particles, the polarization of constraints algebra is
necessarily complex and can be defined by two complementary
(momentum dependent) projection operators. The polarized constraints
are defined as  follows:
\begin{equation}
\label{Gp}
    G^{\alpha}_{(\pm)} = \left (p^{\alpha}_{\;\beta} \pm im(p)
    \delta^{\alpha}_{\;\beta}\right )G^{\beta}\;,
\end{equation}
where $m(p) = \sqrt{-p^2}$ is the mass function. \\ From
(\ref{secondclasscon}) and (\ref{poissonmixed}) it follows that the
systems defined by either $(G^{\alpha}_{(+)},H_{\rm D})$ or
$(G^{\alpha}_{(-)},H_{\rm D})$ are of first class:
\begin{equation}
\label{Palgebra}
    \{\,G^{\alpha}_{(\pm)},G^{\beta}_{(\pm)}\} =
0\;,\;\; \{\,H_{\rm D},G^{\alpha}_{(\pm)}\} = \pm
\frac{ih}{2}m(p)G^{\alpha}_{(\pm)}\;,
\end{equation}
The "classical anomaly" is hidden in the mixed bracket:
\begin{equation}
\label{anom} \{\,G^{\alpha}_{(+)},G^{\beta}_{(-)}\} =
-4im(p)(p^{\alpha \beta} + im(p)C^{\alpha \beta})\;.
\end{equation}
\\
There are at least three good reasons to introduce  the complex Weyl
parametrization of spinor variables now. First of all, the algebra
of functions on the phase space got already complexified. Secondly,
the Weyl spinors constitute the minimal building blocks for
construction of all $\mathbf{SL}(2;\mathbb{C})$ representations. The
last reason is that in these variables the independent constraints
defined by the polarizing projections (\ref{Gp}) are
transparently visible.\\
    The real space of Majorana spinors $(\eta^{\alpha}, \pi^{\alpha})$
    decomposes into, mutually complex
conjugated\footnote{According to common convention $z^{\bar{A}} =
\bar{z}^A$.}, Weyl components $(z^A,\mathfrak{z}^A )_{A=1,2}$ and
$(z^{\bar{A}},\mathfrak{z}^{\bar{A}}) _{\bar{A} = 1,2}\;$. They span
the eigensubspaces of $\gamma^5 = \gamma^0\gamma^1\gamma^2\gamma^3$
Clifford algebra element corresponding to $\pm i$ eigenvalues. This
decomposition is obviously invariant under even subalgebra of
${\mathcal C} (3,1)$ i.e. also under action
$\mathbf{SL}(2;\mathbb{C})$ group. \\
According to (\ref{poissoncan}) the Poisson brackets of the
canonical Weyl variables are given as follows:
\begin{equation}
\label{poissoncomplex}
    \{\mathfrak{z}^A,z^B\} =
\epsilon^{AB}\;,\;\;\{\mathfrak{z}^{\bar{A}},z^{\bar{B}}\} =
\epsilon^{{\bar{A}}{\bar{B}}}\;,
\end{equation}
where $\epsilon^{AB}$ and $\epsilon^{\bar{A}\bar{B}}$ are the matrix
elements of the bilinear form (\ref{convention}) in the complex
basis.
\\
    The second class constraints of (\ref{secondclasscon}) relate the Weyl coordinates:
    $G^A = \mathfrak{z}^{{A}} + z^A = 0$ and $ G^{\bar{A}} =
\mathfrak{z}^{\bar{A}} + z^{\bar{A}}=0$.  Their polarized
counterparts  (\ref{Gp}) can be reexpressed in the following way:
\begin{equation}
\label{GWp}
    G^{A}_{(\pm)} = p^A_{\;\bar{B}}G^{\bar{B}} \pm im(p)
G^A\;\;, \;\;G^{\bar{A}}_{(\pm)} = p^{\bar{A}}_{\;{B}}G^{{B}} \pm
im(p) G^{\bar{A}}\;,
\end{equation}
where $p^A_{\;\bar{B}}$ and $p^{\bar{A}}_{\;{B}}$ are (mutually
complex adjoint) matrix elements of the real operator
$p^{\mu}\gamma_{\mu}$ in the complex basis of Weyl spinors. The
Clifford algebra relations imply that they do satisfy:
$p^A_{\;\bar{B}}p^{\bar{B}}_{\; {C}} = p^2\delta^A_C\,$ and
$p^{\bar{A}}_{\;{B}}p^{{B}}_{\; \bar{C}} =
p^2\delta^{\bar{A}}_{\bar{C}}\,$ .
\\
The Hamiltonian constraint rewritten in terms of Weyl variables
takes the form:
\begin{equation}
\label{kinetic}
    H_{D} = \frac{1}{2}(p^2 + m_0^2) -
\frac{h}{2}(\mathfrak{z}^Ap_{A\bar{B}}z^{\bar{B}} +
\mathfrak{z}^{\bar{A}}p_{\bar{A}B}z^B)\;.
\end{equation}
The Poisson algebra of the complex constraints can be easily
calculated. From (\ref{Palgebra}) it follows that:
\begin{equation}
\label{zero}
    \{G^{A}_{(\pm)},G^{B}_{(\pm)}\} = 0 =
\{G^{\bar{A}}_{(\pm)},G^{\bar{B}}_{(\pm)}\}\;.
\end{equation}
One may check that the functions (\ref{GWp}) are, under the Poisson
bracket, the mass-weighted eigenfunctions of (\ref{kinetic}):
\begin{equation}
\label{masseigen} \{H_D,G^{A}_{(\pm)}\} = \pm
\frac{ih}{2}m(p)G^{A}_{(\pm)}\;,\;\; \{H_D,G^{\bar{A}}_{(\pm)}\} =
\pm \frac{ih}{2}m(p)G^{\bar{A}}_{(\pm)}\;.
\end{equation}
It is not difficult to notice that  $G^{A}_{(\pm)}$ and
$G^{\bar{A}}_{(\pm)}$ are not independent. One finds the following
relation:
\begin{equation}
\label{notindependent}
   G^{\bar{A}}_{(\pm)} = \mp\frac{i}{m(p)}p^{\bar{A}}_{\,{B}}G^{{B}}_{(\pm)} \;.
\end{equation}
From (\ref{zero}-\ref{masseigen}), the conjugation properties
$\overline{G^{A}_{(\pm)}} = G^{\bar{A}}_{(\mp)}$ and the relation
above, it follows, that the systems $(H_D,G^{A}_{(\pm)})$
constitute, mutually complex conjugated, polarized Poisson algebras
of first class.

\section{The quantum model}
The classical system is canonically quantized in the representation
on the space of square integrable functions of the momentum
variables $(p^\mu)$ and  Weyl spinor coordinates
$(z^A,z^{\bar{A}})$. According to (\ref{poissoncan}),
(\ref{poissoncomplex}) and the standard correspondence rules the
canonically conjugated variables are realized as differential
operators: $ x^{\mu} \rightarrow -i{\partial}/{\partial
p_{\mu}}\;\;$ and $\; \mathfrak{z}^A \rightarrow
i\epsilon^{AB}{\partial}/{\partial z^B}$,$\;\mathfrak{z}^{\bar{A}}
\rightarrow i\epsilon^{\bar{A}\bar{B}}{\partial}/{\partial
z^{\bar{B}}}\; $. Under this substitution the constraints of
(\ref{GWp}) take  the following form:
\begin{equation}
\label{qconstraints}
    G^{A}_{(\pm)}= ip^{A {\bar{B}}}\frac{\partial}{\partial
    z^{\bar{B}}}
    \mp m(p)\epsilon^{AB}\frac{\partial}{\partial z^B} +
    p^A_{\;\;\bar{B}}z^{\bar{B}} \pm im(p) z^{A}\;,
\end{equation}
while the canonical hamiltonian (\ref{kinetic}) is transformed into:
\begin{equation}
\label{qham}
     H_{D} = \frac{1}{2}(p^2 + m_0^2) + S\;,\;\;\; {\rm where}\;\;\;
    S = -\frac{ih}{2}({z}^{\bar{B}}p_{\bar{B}}^{\;\;\;A}\frac{\partial}{\partial z^{A}}
    + {z}^{{B}}p_{B}^{\;\;\;\bar{A}}\frac{\partial}{\partial z^{\bar{A}}})\;.
\end{equation}
   As it will be made evident the  operator $S$ above is responsible for spin-mass
    coupling.\\
The generators of $\mathbf{SL}(2;\mathbb{C})$ group are obtained as
the operator counterparts of the conserved classical quantities
corresponding to Lorentz invariance of (\ref{action}):
\begin{equation}
\label{lorentz}
    L^{\mu\nu} = i\left(p^\mu \frac{\partial}{\partial p^\nu} - p^\nu \frac{\partial}{\partial p^\mu}
    \right)+ \frac{i}{2}\left( z^A \sigma ^{(\mu\nu) B}_{\;A} \frac{\partial}{\partial z^B} +
    z^{\bar{A}} \sigma ^{(\mu\nu) {\bar{B}}}_{\;{\bar{A}}} \frac{\partial}{\partial
    z^{\bar{B}}}\right)\;.
\end{equation}
\\
The momenta of the particles were already at the classical level
restricted to the massive region $p^2<0$. This open domain consists
of two disjoint components: the interiors of the future pointed
$p^0>0$ and past pointed $p^0<0$ light cones. The wave functions
with supports in these disjoint regions should be interpreted as
particle and anti-particle states respectively. Hence, the space of
states of the system under consideration decomposes into the direct
sum of two orthogonal subspaces:
\begin{equation}
\label{sum}
    H = H^{\uparrow} \oplus H^{\downarrow}\;,
\end{equation}
consisting of the wave functions with supports in $p^0>0$ and
$p^0<0$ cone interiors. \\
    The physical subspace $H_{\rm{phys}}$
of $H$ should  also be  searched for in the form of the direct sum
corresponding to (\ref{sum}).  The direct summands should be defined
by:
\begin{equation}
    H^{\uparrow\downarrow}_{(\pm)} = \{\Psi_{\pm} \in
H^{\uparrow\downarrow}\;;\;\; G^{A}_{(\pm)}\Psi_{\pm} = 0 =
H_D\Psi_{\pm}\}\;,
\end{equation}
where (without any correlation with $^{\uparrow\downarrow}$ at the
moment) either $G^{A}_{(+)}$ or $G^{A}_{(-)}$ constraints are
imposed.\\
\\
    From the representation theory of the Poincare group
it clearly follows \cite{Lop} that one should look for the solutions
of the constraints equations within the set of functions of the
form:
\begin{equation}
\label{states}
    \Psi_{\pm}(p,z,\bar{z}) = W(z,\bar{z})\Omega_{\pm} (p) \;,
\end{equation}
where $W(z,\bar{z})$ are the polynomials of Weyl variables with
square integrable $p$-dependent coefficients, and $\Omega_{\pm} (p)$
-  the exponential factors of Gaussian type in $(z^A,z^{\bar{A}})$
coordinates. Their presence is essential for the  states (\ref{states}) to be normalizable.\\
   For the  exponential factors
to belong to the physical subspace it is necessary to impose the
constraints equations $G^{A}_{(\pm)} \Omega_{\pm} (p) = 0\;$. Their
unique (up to multiplicative constant) solutions are given by:
\begin{equation}
\label{vacuum}
    \Omega_{\pm} (p) = \exp \pm \frac{z^{\bar{A}}p_{\bar{A}B}z^B}{m(p)}\;.
\end{equation}
According to the convention adopted in (\ref{convention}) the matrix
$(p_{\bar{A}B})$ is negatively defined for $p^0>0$ while it is
positive in $p^0<0$ region. \\
    Consequently, the space of physical states is necessarily of the
following structure:
\begin{equation}
\label{sum1}
     H_{\rm{phys}} = H_{(+)}^\uparrow \oplus
    H_{(-)}^\downarrow\;,
\end{equation}
i.e. the positive frequency physical states are annihilated by
$G^{A}_{(+)}$ and negative frequency physical states occupy the
kernel of $G^{A}_{(-)}$. \\
    From (\ref{lorentz}) and (\ref{vacuum})
it follows that the states $\Omega_{\pm} (p)$ are of scalar
character with respect to $\mathbf{SL}(2;\mathbb{C})$
transformations, i.e. they carry spin zero. For this reason it is
natural to call them the spin vacuum states.\\
    Since the spin vacua (\ref{vacuum}) are in the kernel of the constraints
   (\ref{qconstraints})  their action of on the states (\ref{states}) simplifies
    remarkably:
\begin{equation}
\label{actiononstates}
     G^{A}_{(\pm)}\left( W(z,\bar{z})\Omega_{\pm} (p)\right) =
     \left( D^{A}_{(\pm)}\,W(z,\bar{z})\right)\Omega_{\pm}(p)\;,
\end{equation}
where $D^{A}_{(\pm)}$ denote the differential parts
(\ref{qconstraints}) of $G^{A}_{(\pm)}$.\\
\\
    In order to recover the structure of the space (\ref{sum1}) the
detailed analysis of $H^{\uparrow}_{(+)}$ will be presented here.
The way of  proceeding with $H^{\downarrow}_{(-)}$ is completely
analogous.\\
    Any state from $H^{\uparrow}$ can be represented as a
superposition of the vectors with fixed (common) $(z^A,z^{\bar{A}})$
degree $2j$:
\begin{equation}
\label{degspinstate}
    \Psi_{j}(p,z,\bar{z}) = \sum\limits_{n=0}^{2j} \Psi_{A_{1}\ldots
A_{2j-n}\bar{B}_{1}\ldots\bar{B}_{n}}(p)z^{A_1}\cdots z^{A_{2j-n}}
z^{\bar{B}_1}\cdots z^{\bar{B}_{n}}\Omega_{+} (p) \;.
\end{equation}
The subspace of $H^{\uparrow}$ spanned by the above states is stable
under the action of $\mathbf{SL}(2;\mathbb{C})$ group generators of
(\ref{lorentz}). It contains the positive frequency wave functions
of the particles with spins not exceeding $j$ and is highly
reducible: for example the multiplicity of
 spin $j$ representation in (\ref{degspinstate}) equals to $2j+1$.
\\
This degeneracy is completely removed by the constraints
$G^{A}_{(+)}$: when imposed on the
    states (\ref{degspinstate}) they generate the chain of equations:
\begin{equation}
\label{equation} -(n+1)
p_{A_{2j-n}}^{\bar{B}_{n+1}}\Psi_{A_{1}\ldots
    A_{2j-n-1}\bar{B}_{1}\ldots\bar{B}_{n+1}}(p)=
    i m(p)(2j-n)\Psi_{A_{1}\ldots
    A_{2j-n}\bar{B}_{1}\ldots\bar{B}_{n}}(p)\;,
\end{equation}
where $n = 0,\ldots,2j-1$. These relations can be called the
generalized Dirac equations\footnote{For $j=\frac{1}{2}$
(\ref{equation}) is exactly Dirac equation.} \cite{Lop}. They enable
one to express all fixed $n$ components in the expansion
(\ref{degspinstate}) by the single one. As the root component one
may choose for example the holomorphic part corresponding to $n=0$:
\begin{equation}
\label{holomorphic}
    \Psi_{j}(p, z)=\Psi_{{A}_{1}\ldots
    {A}_{2j}}(p)z^{A_1}\cdots z^{A_{2j}}\Omega_{+} (p)\;.
\end{equation}
Then the recurrence of (\ref{equation}) is solved by:
\begin{equation}
\label{solutionrec}
    \Psi_{A_{1}\ldots A_{2j-n}\bar{B}_{1}\ldots\bar{B}_{n}}(p) =
    \left(\frac{i}{m(p)}\right)^{n}
   \left( \begin{array}{c} 2j \cr n \end{array}\right)
    p_{\bar{B}_n}^{A_{2j-n+1}}\cdots p_{\bar{B}_1}^{A_{2j}}\Psi_{A_1\ldots A_{2j-n+1}\ldots
    A_{2j}}(p)\,.
\end{equation}
 Hence, the constraint equations $G^{A}_{(+)} = 0$ which remove the degeneracy
 from (\ref{degspinstate}) are nothing but spin irreducibility conditions \cite{Lop}.
\\
According to the analysis performed above one is in a position to
introduce the intermediate space of physical off-shell  states. This
space splits into the direct sum:
\begin{equation}
\label{off-shell}
    {\hat H}_{(+)}^\uparrow = \bigoplus_{j\geq 0}{\hat
    H}_{(+)}^{\uparrow\,j}\;,
\end{equation}
where the subspaces ${\hat H}_{(+)}^{\uparrow\,j}$ contain exactly
one family of the particles with fixed spin $j$ but with arbitrary
masses.\\
 In order to recover the physical spectrum one has to impose the hamiltonian
constraint $H_D$ on the spin irreducible states of
(\ref{off-shell}). Luckily, the operator $S$ of (\ref{qham}) is
diagonal on the space of off-shell wave functions from ${\hat
H}_{(+)}^\uparrow$ : $S\Psi_{j}(p,z,\bar{z}) =
-2hjm(p)\Psi_{j}(p,z,\bar{z})$.
    The equation $H_D\Psi_{j}(p,z,\bar{z}) = 0$ imposes the
    following simple
condition on the momentum support:
\begin{equation}
\label{momentum}
    \left( m^2(p)  + 2hj m(p)- m_0^2\right)\Psi_{j}(p,z,\bar{z}) =
    0\;.
\end{equation}
This equation has two real solutions with different signs.  The
positive one is given by\footnote{The negative one has to rejected
as according to (\ref{vacuum}) it would give unormalizable vectors
without physical interpretation as the quantum states.}:
\begin{equation}
\label{m+} m_{j}^{\uparrow} = \sqrt{h^2j^2+m_0^2} - hj \;\;;\;\;
j\geq 0\;.
\end{equation}
In this way the momentum support of $H_{(+)}^\uparrow$ gets reduced
to a single mass-shell corresponding to (\ref{m+}). The reduced
space contains the states of a single particle with fixed spin and
mass.\\
The whole space of physical states $H_{(+)}^\uparrow$ with future
pointed momenta contains the particles with arbitrarily high spins
and with masses tending to zero when their spins grow.\\
In order to summarize the structure of the space of physical states,
it is worth to present the explicit formulae for their scalar
product calculated in terms of the spin root components, chosen in
(\ref{holomorphic}):
\begin{eqnarray}
\label{scalarproduct} \nonumber
    \left(\Psi_i,\Phi_j\right) &=& \cr
    =\delta_{ij}(-1)^{2j} &C_{j}&
     \int \frac{d^4p}{m(p)^{2j}}\; p^{\bar{A}_1 {B}_1 }\cdots p^{\bar{A}_{2j} {B}_{2j}}
    \bar{\Psi}_{\bar{A}_1 \cdots \bar{A}_{2j}}(p){\Phi}_{{B}_1 \cdots
    {B}_{2j}}(p)\theta(p^0)\delta(p^2+m_j^{\uparrow 2})\;,
\end{eqnarray}
with $ C_{j}$ being the positive combinatorial
factor.\footnote{Since the momentum matrices $(p^{\bar{A}B})$ are
negatively defined in $p^0
> 0$ region the presence of $(-1)^{2j}$ guarantees the positivity
of scalar product.}\\
\\
     In the case of the
space  $H^{\downarrow}$ supported by the past pointed momenta one
is, as it was already justified by normalizability arguments, to
impose the complementary $G^{A}_{(-)}$ spin irreducibility
constraints. The analysis analogous to the one performed above leads
to the recurrence formula of the type of (\ref{equation}), and again
gives the
representation of the fixed spins in the irreducible way. \\
    The kinematic constraint (\ref{qham}), applied to spin irreducible states
$\Psi_{j}(p,z,\bar{z})$ with the support on the past pointed
momentum cone, amounts to the following  condition this time:
\begin{equation}
\label{momentum-}
    \left(m^2(p) - 2h j m(p)- m_0^2\right)\Psi_{j}(p,z,\bar{z}) = 0\;,
\end{equation}
which has the unique positive mass solution given by:
\begin{equation}
\label{m-} m_{j}^{\downarrow} = \sqrt{h^2j^2+m_0^2} + hj \;\;;\;\;
j\geq 0\;.
\end{equation}
In contrast to the previous situation the masses of the particles
grow with their spins. \\
\\
The content of the quantum system under consideration can be
summarized as follows. First of all, the model describes the
infinite family of particles with spin. In both, particle ($p^0>0$)
and anti-particle ($p^0<0$) sectors, every spin is represented in
the irreducible way i.e. with multiplicity one.
\\
According to (\ref{m+}) and (\ref{m-}) the masses of particles and
their potential anti-particles are located on two different Regge
trajectories (Fig.1).\\
The mass difference grows linearly with spin:
\begin{equation}
\label{grow} \Delta m_j = m_j^\downarrow - m_j^\uparrow =
2hj\;\;;\;\;j\geq 0\;,
\end{equation}
and for this reason it is justified to call the particles and
anti-particles as being orphaned.
\begin{figure}[h]
\begin{center}
\includegraphics[width=11.5cm]{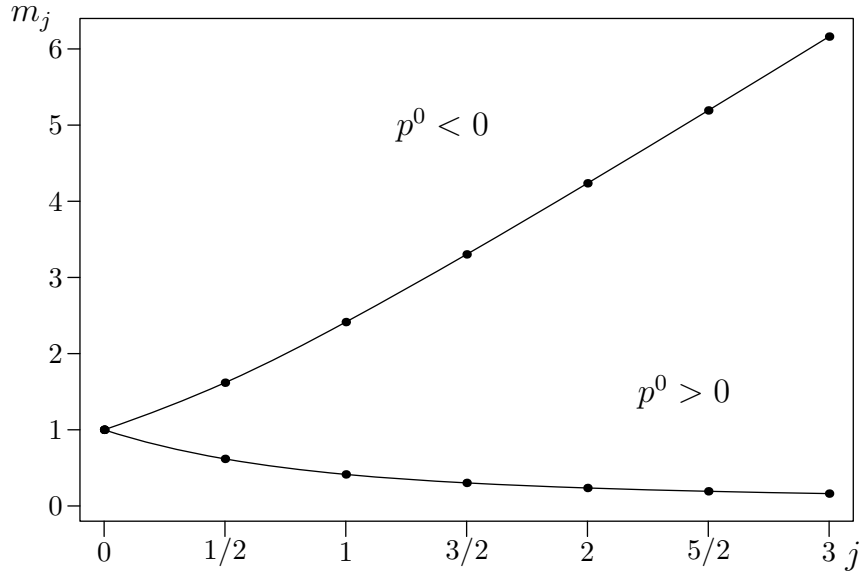}
\end{center}
\caption{The mass spectrum} \label{fig.1}
\end{figure}\\

\section*{Conclusions and outlook}
As it was shown, the simple classical model  considered in this
paper describes, after quantization, the families of particles and
anti-particles located at diverging Regge trajectories with common beginning at spin vacuum.\\
The kinematic equations (\ref{momentum}) and (\ref{momentum-})
besides of the solutions given in (\ref{m+}) and (\ref{m-}) admit
also the solutions with negative masses. The absolute values of
these
masses complete the spectrum by missing CPT related elements.\\
Unfortunately, as already mentioned, they had to be rejected  as
unphysical. One could try to interpret them as negative frequencies
in the rest frame of the particles. It is however excluded by the
obvious reasons: the functions of (\ref{states}) become not
normalizable, and, in fact, they do not belong to the Hilbert space
of
the quantum model. \\
Hence, it seems that the phenomenon of CPT symmetry breaking is the
intrinsic property of the considered system.\\
\\
As it was mentioned in the Introduction, it is possible to try to
restore this symmetry in conceptually simple way - by supplementing
the system by an additional spinorial degree of freedom with
opposite spin-mass coupling. The Lagrange function of (\ref{action})
gets then modified to:
$$
{\cal L}=\frac{1}{2}e^{-1}\dot{x}^2-\frac{1}{2}e
    m_{0}^{2} +\bar{\eta}{\dot \eta} -\frac{h}{2} \dot{x}\cdot j_\eta +
    \bar{\zeta}{\dot \zeta} +\frac{h}{2} \dot{x}\cdot j_\zeta\;.
$$
This simple modification leads however to an additional quartic term
in Dirac hamiltonian, which describes the cross-interaction of
spinor currents. For this reason, the analysis of the model extended
in this way is much more difficult, and is postponed to the future publication.\\
\\
One more remark is in order here. From (\ref{m+}) it is evident that
the model (at least in the case of $m_0 = 0$) admits massless
solutions. One would like to obtain these states by some limiting
procedure out of the massive ones. This procedure is not
straightforward as the spin vacuum states of (\ref{vacuum}) do
vanish when the mass tends to zero. For this reason, the massless
limit has to be defined in some more subtle way, which would in
addition give as an outcome, the one component wave functions for
the
massless particles. This problem is left open. \\
\\
It is worth to mention finally, that  (after the
particle-antiparticle symmetry is restored) the local field theory
based on the system of the type considered here, can be used as a
starting point to the analysis of the Friedmann-type cosmological
model with multi-spin sources (see e.g. \cite{current} and the
references therein).
The work in this direction already started.\\

\section*{ Acknowledgements}
The authors would like to thank prof. J.Lukierski for advices and
reading of the manuscript. \\
One of the authors (M.D.) would like to thank the friends and
colleagues from the Institute of Theoretical Physics in Bia{\l}ystok
for warm hospitality. He would also like to thank dr. S.Ciechanowicz
for discussion. Finally, M.D. would like to thank his colleagues
M.Kucab, T.Nowak, B.\.{Z}ak and W.\.{Z}ak for their inspirations
during the training sessions. Special thanks are
due to Zygmunt Gosiewski. \\
This work is partially supported by KBN grant 1P03B01828.


\begin{thebibliography}{99}
\bibitem{HDS} Z.Hasiewicz, F.Defever, P.Siemion,
Int. J. Mod. Phys. A 7 (1992) 3979-3996
\bibitem{GT} V.D.Gershun, V.I.Tkach, JETP Lett. (1979) 320
\bibitem{HPPT} P.S.Howe, S.Penati, M.Pernici, P.Townsend, Phys. Lett.
B215 (1988) 255
\bibitem{A2} P.P.Srivastava, Nuovo Cimento Lett. 19 (1977) 239
\bibitem{A3} A.Barducci, L.Lusanna, J. Phys. A16 (1983) 1993
\bibitem{A1} A.Bette, J.de Azcarraga, J.Lukierski, C.Miquel-Espanya,
Phys. Lett. B595 (2004)
\bibitem{LF} S.Fedoruk, J.Lukierski, Phys. Lett. B632 (2006)
371-378
\bibitem{LF1} S.Fedoruk, J.Lukierski, {\it Higher spin particles with bosonic
counterpart of supersymmetry} - Workshop on Supersymmetries and
Quantum Symmetries (SQS'05), Dubna, Russia, 27-31 Jul 2005
\bibitem{andrzej} A.Frydryszak, {\it Lagrangian models of particles
with spin}, Published in {\it From field theory to quantum groups},
Singapore World Scientific Publishing (1996)
\bibitem{daszhasjas1} M.Daszkiewicz, Z.Hasiewicz, Z.Jaskolski, Nucl. Phys. B514 (1998) 437-459
\bibitem{daszhasjas2} M.Daszkiewicz, Z.Hasiewicz, Z.Jaskolski, Phys. Lett. B454 (1999) 249-258
\bibitem{Pol} J.Polchinski, {\it String theory: An introduction to the bosonic
string}, Cambridge, UK: Univ. Pr. (1998) 402 p.
\bibitem{field}
A.R.Swift, Phys. Rev. 176 (1968) 1848-1855 \\
W.J.Zakrzewski, Nuovo Cim. A60 (1969) 263-290 \\
A.R.Swift, R.W.Tucker, Nuovo Cim. A67 (1970) 345-355 \\
A.R.Swift, R.W.Tucker, Phys. Rev. D1 (1970) 2894-2900
\bibitem{plu}
M.Plyushchay, Mod.Phys.Lett. A3, 1299 (1988) \\
M.Plyushchay, Int. J. Mod. Phys. A4 3851 (1989)
\bibitem{schroedinger} E.Schroedinger, Sitzugsh. Preuss. Akad. Wiss.
Phys.-Math. Kl. 24 (1930) 418; 3 (1931) 1
\bibitem{BB} A.O.Barut, A.J.Bracken, Phys. Rev. D23 (1981) 2454
\bibitem{rivas} M.Rivas, {\it Classical elementary particles, spin, zitterbewegung and all that} -
physics/0312107
\bibitem{positionoperators1} T.D.Newton, E.P.Wigner, Rev. Mod.
Phys. 21 (1949) 400
\bibitem{positionoperators2} A.S.Wightman, Rev. Mod.
Phys. 34 (1962) 845
\bibitem{positionoperators3} G.W.Mackey, Bull. Am. Math. Soc. 69
(1963) 628
\bibitem{dirac} P.A.M.Dirac, {\it Theory of electrons and positrons}
Nobel Lecture, December 12, 1933, from Nobel Lectures, Physics
1922-1941, Elsevier Publishing Company, Amsterdam (1965)
\bibitem{GB} S.Gupta, Proc. Roy. Soc. A63 (1950) 681 \\
K.Bleuler, Helv. Phys. Acta 23 (1950) 567
\bibitem{Lop}
Jan {\L}opusza\'{n}ski {\it Spinor Calculus} PWN Warsaw 1985 (in Polish)\\
A.O.Barut, R.R\c{a}czka {\it Theory of Group Representations and
Applications} PWN Warsaw 1977
\bibitem{Jost}R.F.Streater, A.S.Wightman {\it PCT Spin - Statistic
and all that} Benjamin, New York 1964 \\
R.Jost {\it The General Theory of Quantized Fields} AMS, Providence,
Rhode Island 1965
\bibitem{brst} I.A.Batalin, E.S.Fradkin, Nucl. Phys. B279 (1987)
\bibitem{DHW} M.Daszkiewicz, Z.Hasiewicz, C.Walczyk, Rep. Math. Phys. Vol. 59 (2007) 2 185-208
\bibitem{Eva} B.Taylor, C.Chambers, W.Hiscock,
Phys. Rev. D58 (1998) 044012
\bibitem{Schwarzschild}
D.Boulware, Phys. Rev. D12 (1975) 350
\bibitem{current} M.O.Ribas, F.P.Devecchi, G.M.Kremer, Phys. Rev. D72 (2005) 123502


\end{thebibliography}
\end{document}